\documentclass[fleqn,usenatbib]{mnras}
\usepackage{amsmath}
\usepackage{graphicx}
\usepackage{grffile}
\usepackage{hyperref}
\usepackage{txfonts}
\usepackage{natbib}
\usepackage{bm}
\usepackage{calc}
\usepackage{color}

\usepackage[T1]{fontenc}
\usepackage{ae,aecompl}

\title[New changing look case in NGC~1566  ] {New changing look case in NGC 1566}

\author[V. L. Oknyansky et al.]{V. L. Oknyansky,$^{1}$\thanks{E-mail: oknyan@mail.ru}
H. Winkler,$^{2}$
S. S. Tsygankov,$^{3,4}$
V. M. Lipunov,$^{1}$
\newauthor E. S. Gorbovskoy,$^{1}$
F. van Wyk,$^{2}$
D. A. H. Buckley,$^{5}$
N. V. Tyurina$^{1}$
\\
\\
$^{1}$ M. V. Lomonosov Moscow State University, Sternberg Astronomical Institute, 119234, Moscow, Universitetsky pr-t, 13, Russian Federation\\
$^{2}$ Dept. Physics, University of Johannesburg, PO Box 524, 2006 Auckland Park, South Africa\\
$^{3}$ Department of Physics and Astronomy, FI-20014 University of Turku, Finland \\
$^{4}$ Space Research Institute of the Russian Academy of Sciences, Profsoyuznaya Str. 84/32, Moscow 117997, Russia \\
$^{5}$ The South African Astronomical Observatory, PO Box 9, 7935 Observatory, South Africa}
\date{Accepted XXX. Received YYY; in original form ZZZ}

\pubyear{2018}

\begin{document}
\date{Received ... Accepted ...}
\pagerange{\pageref{firstpage}--\pageref{lastpage}}
\maketitle{}

\label{firstpage}

\begin{abstract}
 We present a study of optical, UV and X-ray light curves of the nearby changing look active galactic nucleus in the galaxy NGC 1566 obtained with the {\it Neil Gehrels Swift Observatory} and the MASTER Global Robotic Network over the period 2007 - 2018. We also report on our optical spectroscopy at the South African Astronomical Observatory with the 1.9-m telescope on the night 2018 August 2--3.  A substantial increase in X-ray flux by 1.5 orders of magnitude was observed following the brightening in the UV and optical bands during the last year. After a maximum was reached at the beginning of 2018  July the fluxes in all bands decreased with some fluctuations. The amplitude of the flux variability is strongest in the X-ray band and decreases with increasing wavelength. Low-resolution spectra reveal a dramatic strengthening of the broad emission as well as high-ionization [FeX]6374\AA ~lines. These lines were not detected so strongly in the past published spectra.  The change in the type of the optical spectrum was accompanied by a significant change in the X-ray spectrum. All these facts confirm NGC~1566 to be a changing look Seyfert galaxy.

\end{abstract}




\section{Introduction}

The Changing Look active galactic nuclei (CL AGNs) are objects which undergo dramatic variability of the emission line profiles and classification type, which can change from  type 1 (showing both broad and narrow lines) to type 1.9 (where the broad lines almost disappear) or vice versa within a short time interval (typically a few months). The most extreme examples of known CL AGNs exhibit very strong changes not only in their broad Balmer lines, but also in their more narrow high-ionization lines like [FeVII], [FeX] and [FeXIV] \citep[see e.g., ][]{Parker2016}. There are currently only a few dozen cases of CL AGNs recorded, but their number is growing steadily all the time, in particular during last 5 years \citep[see e.g., ][],{Shappee2014, Koay2016, MacLeod2016, Rinco2016, Rumbaugh2018, Yang2018}. More than 20 AGNs have been witnessed as changing their spectral look in the X-ray domain \citep[see references in][]{Ricci2016}. The origin of the UV and optical variability in AGN, and its correlation with the X-ray variability, is not well understood. Therefore investigations of CL AGNs during such transitions can be very helpful for understanding the central structure of AGNs as well as the physics underpinning such dramatic changes.

NGC~1566 is a galaxy with a very well-studied variable active nucleus. It is a nearly face-on spiral galaxy with morphological type SAB(s)bc. It is one of the brightest ($V\approx$10.0 mag) and nearest galaxies with AGN in the South Hemisphere (the distance is still subject to large uncertainty and so we adopt the  median value given by NED{\footnote{The NASA/IPAC Extragalactic Database (NED) is operated by the Jet Propulsion Laboratory, California Institute of Technology, under contract with the National Aeronautics and Space Administration}} , i.e. $D\approx7.2$ Mpc). The NGC~1566 nucleus exhibits a Seyfert emission spectrum \citep{Vaucouleurs1961, Shobbrook1966} that was later classified as type 1. Spectral variations had been witnessed in this object even in some of the earliest spectroscopic investigations in 1956 \citep{Vaucouleurs1961} and in 1966 \citep{Shobbrook1966, Pastoriza1970}. Other early observed spectra \citep{Osmer1974} showed that the H$\alpha$ and H$\beta$ profiles had broad components with strong asymmetry towards the red. NGC~1566 can be classified as a CL object since it has exhibited strong variations in the past decades \citep{Alloin1985, Kriss1991, Baribaud1992, Winkler1992}. During the 1970s and 1980s the object spent much time in a low-luminosity state where the broad permitted lines were nearly undetectable \citep{Alloin1986}. However it had several recurrent brightening events when its type changed from Sy 1.9-Sy 1.8 to Sy 1.5-Sy 1.2 states. The most significant recorded past outbursts were observed in 1966 and 1992 \citep{Silva2017}. The first multiwavelength investigations of the NGC~1566 variability from X-ray to IR, as well as first IR reverberation mapping in the object, were done by \cite{Baribaud1992}. They found a possible delay in the variations of the IR $K$ band relative to those of H$_\alpha$ of about $2\pm1$ month. The lag of the $K$ band variability relative to $J$ was however elsewhere found to be less than 20 days \citep{Oknyansky2001}. \cite{Alloin1985} and \cite{Silva2017} found the broad line region upper limit radius to be 20 light-days and 15 light-days respectively. These values together with the IR time delays may be used to estimate the location of the inner edge of the dust torus \citep[see e.g.,][]{Clavel2000, Netzer2015}. In the X-ray regime, studies of the Fe$\alpha$ emission line suggest that it is linked to reflection from the torus \citep{Kawamuro2013}. 

On 2018 June 12-19 data from the {\it INTEGRAL} observatory showed that NGC~1566 was in outburst in hard X-rays \citep{Ducci2018}, and led to follow-up observations with the {\it Swift} observatory \citep{Ferrigno2018, Grupe2018, Kuin2018}. An ASAS-SN $V$-band light curve for the period 2014--2018 shows that the nuclear started to brighten significantly around 2017 September  \citep{Dai2018}. The 2014-2018 mid-infrared light curves of the object show that the nucleus brightened by 1 mag at 3.4 $\mu$m and 1.4 mag at 4.6 $\mu$m between 2017 January  and 2018 July \citep{Cutri2018}.      

In this paper we report on follow-up optical and UV photometric observations (MASTER (Mobile Astronomical System of TElescope Robots); {\it Swift}/UVOT), as well as new X-ray ({\it Swift}/XRT) data. We also report on our optical spectroscopy at the South African Astronomical Observatory (SAAO) of what is one of the brightest and most nearby CL cases witnessed to date \citep{Oknyansky2018B}.

\section{Observations, instruments and reduction}

\subsection{{\it Swift}: optical, ultraviolet, and X-ray observations}

\begin{figure}
\includegraphics[scale=0.8,angle=0,trim=0 0 0 0]{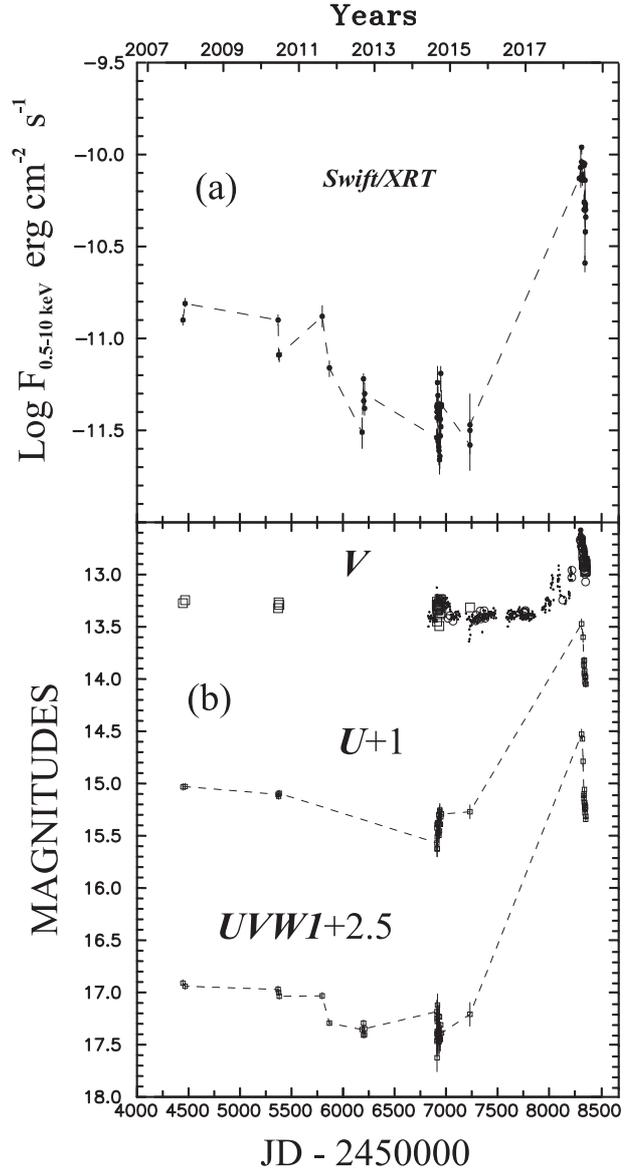}
 \caption {Multiwavelength observations of NGC~1566 spanning from 2007 Dec. 11 through 2018 Aug. 25. {\it Top panel:} The {\it Swift}/XRT 0.5--10 keV  X-ray flux (in erg cm$^{-2}$ s$^{-1}$) -- (filled circles). {\it Bottom panel:} Optical--UV photometric observations. The large open circles are  MASTER unfiltered optical photometry of NGC~1566 reduced to the $V$ system while the points are $V$ ASAS-SN (nightly means) reduced to the {\it Swift} $V$ system. The open boxes are $V$ data obtained by {\it Swift}. The filed circles are MASTER $V$ photometry. The small open boxes are $U$ and $UVW1$ UVOT {\it Swift} photometry.}
    \label{fig1}
\end{figure}

\begin{figure}
\includegraphics[scale=0.55,angle=0,trim=0 0 0 0]{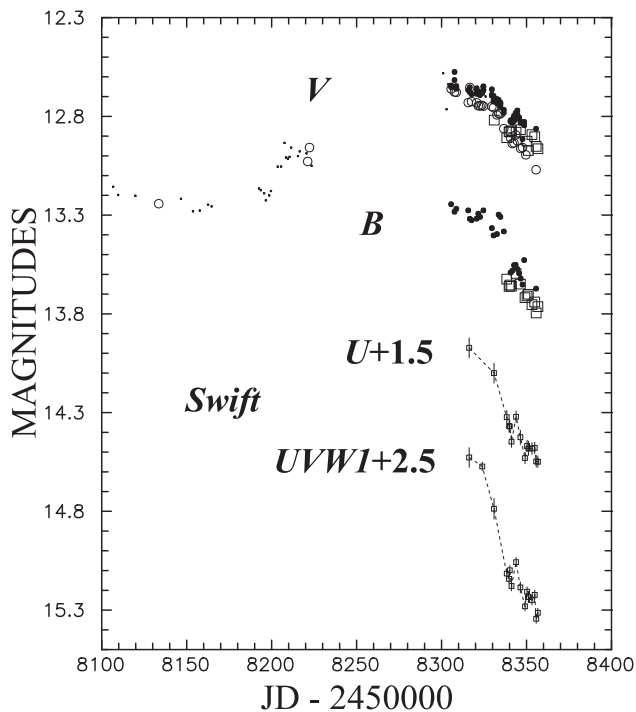}
 \caption{Optical--UV photometric observations of NGC~1566 shown for the last year only. The large open circles are MASTER unfiltered optical photometry of NGC~1566 reduced to the $V$ system while the points are $V$ ASAS-SN (nightly means) reduced to the {\it Swift} $V$ system. The open boxes are $BV$ data obtained by {\it Swift}. The filed circles are MASTER $BV$ photometry. The small open boxes are $U$ and $UVW1$ UVOT {\it Swift} photometry.}
    \label{fig2}
\end{figure}

\begin{figure}
\includegraphics[scale=0.45,angle=0, bb = 40 255 555 700]{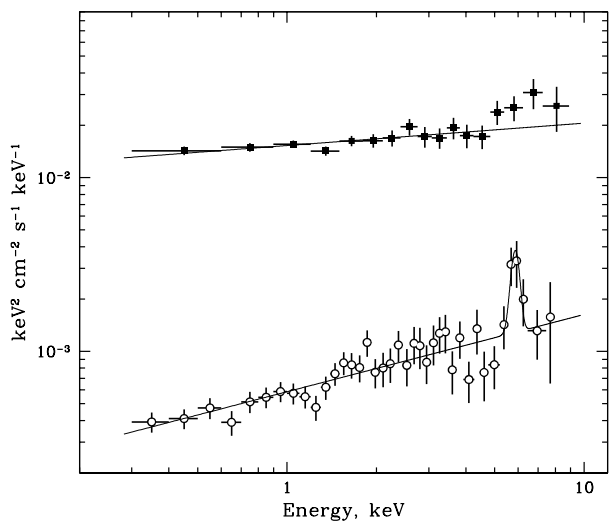}
 \caption {Energy spectra of NGC~1566 obtained with the {\it Swift}/XRT telescope in low state (MJD 56185--56210; open circles) and high state (MJD 58339.7;  squares). Solid lines correspond to the best-fitting models consisting of the power law in the high state and power-law plus Gaussian emission components in the low state.}
    \label{fig3}
\end{figure}

The {\it Neil Gehrels Swift Observatory} \citep{Gehrels2004} has monitored NGC~1566 regularly for years starting from 2007. Some results from these observations have been previously published \citep{Kawamuro2013, Ferrigno2018, Grupe2018, Kuin2018}. In the current work all the data (both from the XRT and UVOT telescopes) were re-reduced uniformly \citep[the same way as in ][]{Oknyansky2017A}, enabling us to trace the evolution of the behaviour of the source on a longer time-scale.

The XRT telescope \mbox{\citep{Burrows2005}} was operating both in photon counting and windowed timing  modes depending on the brightness of NGC~1566. The spectra were reduced using the standard online tools provided by the UK {\it Swift} Science Data Centre \citep[\url{http://www.swift.ac.uk/user_objects/};][]{Evans2009}. Taking into account low count statistics we binned the spectra in the 0.5--10 keV range to have at least one count per energy bin and fitted them using the W-statistic \citep{Wachter1979}.\footnote{See {\sc xspec} manual; \url{https://heasarc.gsfc.nasa.gov/xanadu/} \url{xspec/manual/XSappendixStatistics.html}}
To get the source flux in physical units we fitted all the spectra with a simple absorbed power-law model. The absorption parameter was free to vary, however we did not detect any significant absorption in the data either in high or in low flux states. The resulting light curve in the 0.5--10 keV band is shown in Fig.~\ref{fig1}(a).

The {\it Swift} Ultraviolet/Optical Telescope (UVOT) observes in different bands ($V$, $B$, $U$, $UVW1$, $UVW2$, $UVM2$) simultaneously with the XRT telescope, thus making it possible to get a broad-band view from the optical to X-rays.
The image analysis has been done following the procedure described on the web-page of the UK {\it Swift} Science Data Centre. Photometry was performed with the {\tt uvotsource} procedure with aperture radii of 5 and 10 arcsec for the source and background, respectively. The background was chosen with the centre about 1 arcmin away from the galaxy for all filters.

The XRT and UVOT observation results are presented in Fig.~\ref{fig1}--\ref{fig3} and will be discussed below  in Section 3.1.

\subsection{Observations with the \uppercase{MASTER} network}

MASTER \citep{Lipunov2010} is a fully automated network of telescopes. The global MASTER robotic net for space monitoring was developed at the M.V. Lomonosov Moscow State University and consists of eight observatories. Here we are using data just from one of these observatories located at SAAO. This MASTER--SAAO observatory contains two 40-cm wide field telescopes (MASTER-IIs) with a combined field of view of 8  deg$^2$. Each MASTER II telescope is equipped with fast frame rate industrial GE4000 CCD cameras from AVT company (former Prosilica), which have a detector format of 4008$\times$2672 pixels and an area of 24$\times$36 mm. The MASTER-II instruments are able to provide surveys at a limiting celestial magnitude of 20 on dark, moonless nights. Observations can be made with Johnson $BVRI$ filters, or without a filter for integrated (white) light. Details of MASTER can be found in \cite{Kornilov2012}. One of the goals of the MASTER is investigations of the transient variability of AGNs. One example is the investigation of the CL AGN NGC~2617 \citep{Oknyansky2017A}. Here we present  the MASTER optical photometry of the CL AGN NGC 1566.
To minimize the differences with the {\it Swift} UVOT observations, we performed photometry using a 5 arcsec radius aperture on all our data. We measured the background within an annulus of radii 35--45 arcsec. The calibration of $BV$ magnitudes was done relative to the comparison stars from
\url{http://www.astro.gsu.edu/STARE/ngc1566.html}. The unfiltered data were reduced to the system of the ASAS-SN \citep[All-Sky Automated Survey for Supernovae;][]{Shappee2014, Kochanek2017, Dai2018} $V$--band  (using 14 common dates of observations), whereafter all these $V$ data were converted the same way to the $V$ UVOT {\it Swift} system.

The MASTER observation results are presented in Fig.~\ref{fig1} and \ref{fig2} and will be discussed below in Section 3.2.

\subsection{ Optical  spectral observations and reductions}

\begin{figure}
	\includegraphics[scale=0.49,angle=0,trim=0 0 0 0]{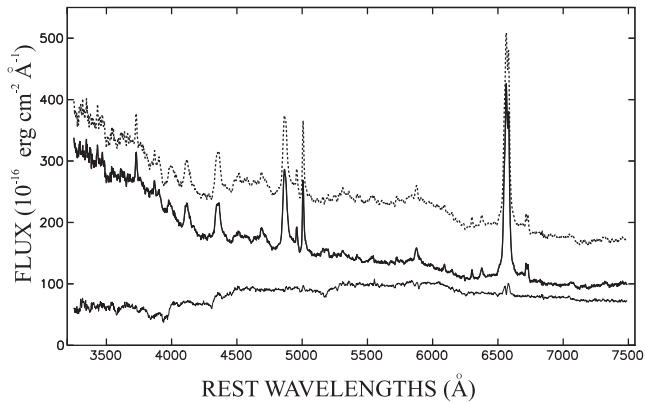}

    \caption{The isolated nuclear non-stellar spectrum (solid line) in NGC~1566 obtained by subtraction of the host galaxy spectrum (thin line) from the original spectrum (dashed line). (See details in the text.)}
    \label{fig4}
\end{figure}

\begin{figure*}
	\includegraphics[scale=0.95,angle=0,trim=0 0 0 0]{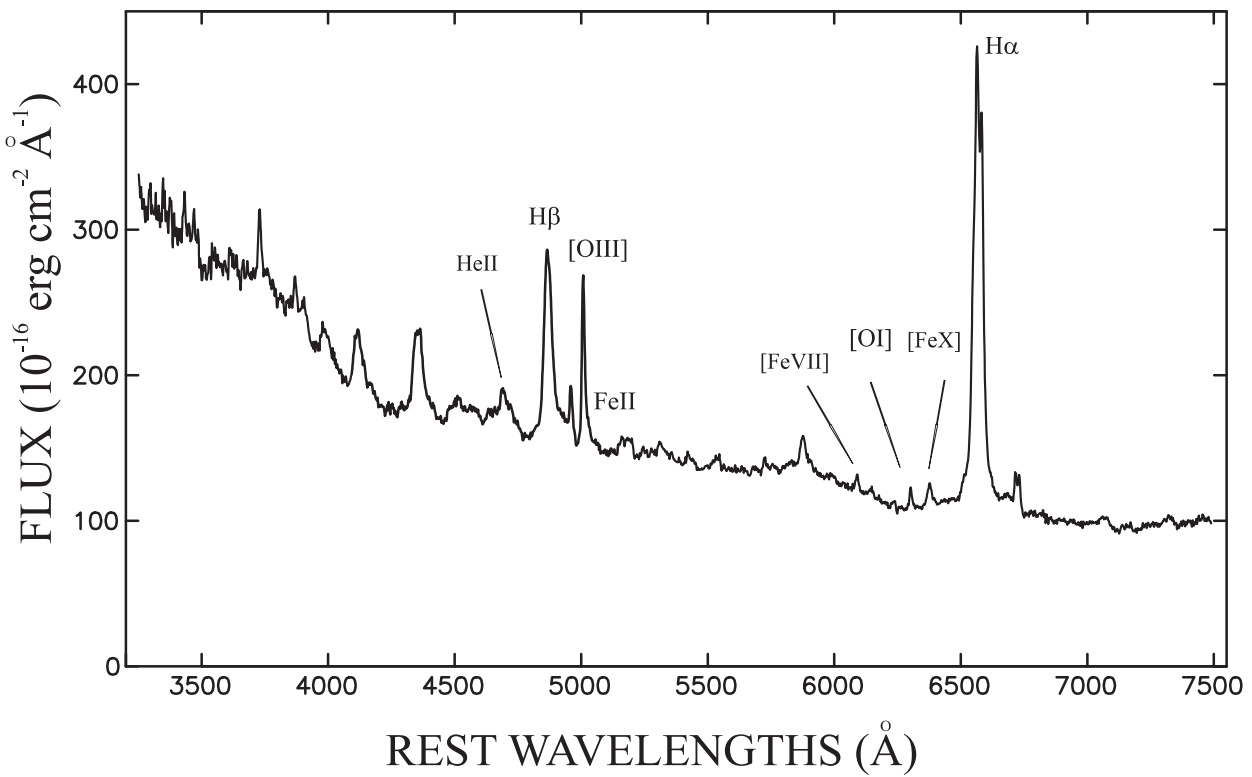}

    \caption{The isolated nuclear non-stellar spectrum in NGC~1566 obtained by subtraction of the stellar spectrum (see details in the text.)}
    \label{fig5}
\end{figure*}

Low-resolution spectra of NGC~1566 were obtained during the night 2018 August 2--3 with the 1.9 m telescope at SAAO in Sutherland. We used the Cassegrain spectrograph  with a 
low-resolution (300 grooves per mm)  grating and a slit width of 2.7 arcsec  to give  spectral range from 3300  to 7500\AA~ and  nominal resolution about 7 \AA.  The spectrograph slit was oriented at a position angle of 90{$^\circ$}. Two spectra of 600 s were taken and were later combined into a single spectrum using an average. The wavelength and flux calibration were achieved by bracketing the AGN spectra with Ar--spectra, and by means of an observation of the spectrophotometric standard star Feige 110 \citep{Hamuy1994}. The spectrum of the nuclear region, displayed as a dashed line in Fig.~\ref{fig4}, represents the flux collected within 5 arcsec east or west of the nuclear position.  This nuclear spectrum includes emission for the nearby HII region \citep{Silva2017}  and contamination from the host galaxy. The HII emission is not relevant to our analysis since these lines are much more narrow  and weaker  \citep[see figs 10--11 in][]{Silva2017} than the AGN narrow components of the emission lines. Their presence just represents a small additional constant contribution to the AGN emission lines which is smaller than the broad line flux uncertainty determined in Section 3.3. The host galaxy contribution to the object spectrum was estimated by scaling the off--nucleus spectrum measured between 6 and 13 arcsec from the nucleus (both to the east and to the west). The host galaxy spectrum therefore avoids an H~II region located approximately 1 arcsec from the nucleus \citep{Silva2017}. This off-nuclear spectrum was then scaled by a factor that was adjusted until the stellar absorption features matched those in the object spectrum. The stellar spectrum is also shown in Fig.~\ref{fig4}, and provides a good match to the Sb galaxy template given by \cite{Kinney1996}. Note that the weak emission near H$\alpha$ is normal in galaxies of this morphological class, which confirms that there is no significant contamination from the nuclear emission lines in the adopted stellar background spectrum. The isolated nuclear nonstellar spectrum  in NGC~1566 obtained by subtraction of the stellar spectrum  from the original spectrum is displayed as a solid line in Fig.~\ref{fig4} and bigger in size in Fig.~\ref{fig5}.  

\section{Resuts}

 \subsection{{\it Swift} XRT and UVOT results}

The resulting light curve (XRT) in the 0.5--10 keV band (spanning from 2007 Dec. 11 to 2018 Aug. 25) is shown in Fig.~\ref{fig1}(a).
Our analysis revealed a strong dependence of the spectral photon index on the source luminosity. This is clearly demonstrated in Fig.~\ref{fig3}, where two spectra are shown for very different intensity states. The low-state spectrum (open circles), collected in 2012  September ({\it Swift}/XRT ObsIds 00045604004-00045604008, where Obs Id is a numeric value that uniquely identifies an observation),  is well fitted by a power-law model with a photon index of $1.54\pm0.06$ and a flux of $F_{\rm 0.5-10 keV}=(4.7\pm0.2)\times10^{-12}$ erg cm$^{-2}$ s$^{-1}$. In addition to the power-law continuum component, the fit required the inclusion of a narrow emission line near 6 keV. This spectrum for the low state is similar to one presented by \cite{Kawamuro2013}. In the bright state ({\it Swift}/XRT Obs Id 00035880015; squares in Fig.~\ref{fig3}) the source flux increased by a factor of $\sim20$ to $F_{\rm 0.5-10 keV}=(8.2\pm0.2)\times10^{-11}$ erg cm$^{-2}$ s$^{-1}$, and the spectrum became significantly softer, with a photon index of $1.88\pm0.03$. The low- and high-state spectra are very different in shape and luminosity, with the soft excess brightening by much more than the hard X-rays. The soft X-ray excess produces most of the ionizing photons, so its dramatic change must lead to a strong outburst of the broad line region, and could therefore be the driver of the changing-look phenomenon. Some excess flux at energies above $\sim5$ keV seen in the bright state may correspond to the iron line or a reflection component \citep[see e.g.,][]{Kawamuro2013}.

Light curves in optical and UV bands ($UV$ and $UVW$) are shown in Fig.~\ref{fig1}(b) for 2007-2018 and in $UBV$ and UV bands  in Fig.~\ref{fig2} for just the last year. Given the very high correlation between variations in different UV bands, we present the light curve just for one of these bands, $UVW1$. 

Analysis of the {\it Swift}/XRT data demonstrates a substantial increase in flux by 1.5 orders of magnitude following the brightening in the UV and optical bands during the past year, with maximum reached during the end of June -- beginning of 2018 July. The minimum recorded level of X-ray flux was about 2.2 $\times$ $10^{-12}$~erg~cm$^{-2}$~s$^{-1}$ (MJD 56936), while the maximum flux measured was 1.1 $\times$ $10^{-10}$~erg~cm$^{-2}$ s$^{-1}$ (MJD 58309), i.e. a ratio of nearly 50 times. Day-to-day fluctuations were also observed a few times on scales much higher than for UV and optical fast variations. The difference is in part due to a lower galaxy contribution in X-rays and can also be explained by the relatively smaller size of the X-ray radiation region. After the maximum , the fluxes in all bands were decreasing quite fast, with some fluctuations. As can be seen in the figures, all variations in the optical, UV and X-ray are correlated. The amplitude is largest for X-rays and decreases with increasing wavelength. Possible lags between these variations will be discussed in future publications.

\subsection{MASTER results in comparison with ASAS-SN and {\it Swift} photometry}

The MASTER photometry  presented here includes old unfiltered archival data from 2014, new intensive monitoring in the $B$ and $V$ bands, as well as unfiltered $W$ band data starting from 2018 July 6 
(Fig.~\ref{fig1}b and Fig.~\ref{fig2}). We show the light curve for  our unfiltered data  reduced  to   the $V$ UVOT {\it Swift} system. The mean nightly (ASAS-SN) $V$-band light curves are also shown in Fig.~\ref{fig1}b and Fig.~\ref{fig2}. The nominal errors for these magnitudes are no more than 0.01 mag in that system, and about three times bigger after reducing these to a smaller aperture. The actual uncertainties may be deduced from the dispersion of the points in the light curve. As is seen from the light curves, our magnitudes are in good agreement with the ASAS-SN and {\it Swift} $V$-band light curves for 2014--2018. The brightening at the beginning of the ASAS-SN 
$V$-band light curve in 2014--2015 is due to supernova ASASSN-14ha (SN), which does not affect our data as our aperture excluded the SN. From 2018 July 6 we started intensive observations in the $B$ and $V$ bands using the MASTER telescopes located in South Africa (SAAO).  We have combined these estimates in Fig.~\ref{fig1}b and Fig.~\ref{fig2}. We present only mean nightly values for the MASTER photometry in all bands. Usually, we obtained two to six observations in each band per night. The nominal errors of the plotted magnitudes are mostly less than the sizes of the shown points.

Our data ({\it Swift} and MASTER) show a roughly linear decline (after the maximum at the beginning of July) that is clearly seen in all bands. Here the average brightness drop rate is about 0.005$\pm$0.001 mag in $V$, 0.009$\pm$0.02 mag in $B$, 0.014$\pm$0.002 mag in $U$ and 0.020$\pm$0.002 mag in $UVW1$ per day, respectively. Fluctuations seen during July--August probably are real since they are visible in all bands. For example, the same local maximum was seen on MJD 58359 at all wavelengths from X-ray to $V$.

\subsection{Optical spectrum: new results}
The spectra in Fig.~\ref{fig5} reveal a dramatic strengthening of the broad emission lines compared to past published ones. This confirms NGC~1566 to be a changing look Seyfert galaxy. 
 
The Balmer lines exhibit a complex profile consisting of (a) a narrow component with the same profile as the nearby [OIII] and [NII] forbidden lines, (b) a broad component with a Gaussian profile and (c) irregularly shaped blue and red wings/humps. In view of the relatively wide slit and low-resolution grating employed, and also because the broad lines are dominant, the Balmer narrow line strengths could not be established to significant accuracy. The hydrogen narrow components were therefore generated by adopting the [OIII]5007\AA-to-H$\beta$ and [NII]6584\AA-to-H$\alpha$ ratios determined by \cite{Silva2017}. The peak wavelengths of the Gaussian broad components ($215\pm30\,km\,s^{-1}$ for H$\alpha$ and $470\pm30$\,km\,s$^{-1}$ for H$\beta$) are slightly lower than was measured by \cite{Silva2017}, while the widths of these components remained consistent with the values determined in that study. Furthermore:

1. H$\beta$ is quite a lot brighter than [OIII]5007\AA. The total H$\beta$ to [OIII]5007\AA line ratio is about 4.2$\pm$0.4, corresponding to a Sy1.2 classification according to the criteria proposed by \cite{Winkler1992}. This ratio is not significantly affected by the narrow component uncertainty because the narrow H$\beta$ intensity is only of the order of 10\% of [OIII]5007\AA. H$\beta$ has not been observed at this strength before in this object.

2. The H$\alpha$ to H$\beta$ ratio for the Gaussian broad components is 2.7$\pm$0.3 . This is consistent with other recent studies \citep[e.g. ][]{Silva2017}, that also found that the obscuration of the broad line region is negligible.

3. The HeII4686\AA ~emission feature is bright and broad. No such strong HeII was visible in the past published spectra \citep[see e.g., ][]{Kriss1991, Winkler1992}.
 
4. The [FeX]6374\AA ~coronal emission line is stronger than [OI]6300\AA, something that has not been seen before in NGC~1566. The ratio is about 1.4$\pm$0.2 after contamination from [OI]6363\AA ~is removed using the assumption that the ratio of [OI]6300\AA~ to [OI]6363\AA ~is 2.997 \citep{Storey2000}. One previous measurement of the coronal line, in a spectrum from 1988 January, recorded this ratio as 0.43 \citep{Winkler1992}. Furthermore, [FeVII]6086\AA ~is quite prominent in the spectrum, and is also stronger than previously recorded. The variability of high-ionization coronal lines is not surprising since these would be expected to arise in the very inner part of the NLR (see e.g. discussion by \citep{Peterson1988} and \citep{Ulrich1997}) or at the inner torus wall \citep{Rose2015}. It was noted that the [FeX]6374\AA ~in some AGNs is broader than for other forbidden lines \citep{Netzer1974, Osterbrock1982}. The variability of coronal lines has also been detected in a number of CL AGNs: NGC~4151, NGC~5548, NGC~7469, 3C~390.3 and others \citep{Oknyansky1982, Veilleux1988, Oknyansky1991, Landt2015, Landt2015b, Parker2016}.

5. A strong UV continuum is clearly seen in our spectrum, and was far more prominent than what is visible in spectra collected during earlier low states.

6. FeII emission is evidently much stronger now than in recent years. For example, the multiplet 42 lines of FeII are clearly seen superimposed on the [OIII] lines in our spectrum, but these were not detected in NGC~1566's low state \citep[see fig. 10 in ][]{Silva2017}.

More details will be presented in a future publication where we will examine the spectral development and profile variations in additional spectra arranged to be collected in the coming season.

\section{Discussion}

NGC~1566 is one of the clearest cases of Seyfert spectra ranging from type 1.2 to type 1.9 AGNs, all being confirmed in the same object at different epochs. In view of the galaxy's relative proximity and brightness, it also offers one of the best opportunities for studying this phenomenon. What must happen to make such a dramatic change possible? CL AGNs like NGC~1566 present problems for the simplest unification models in which type 2 and type 1 AGNs arise under the same processes, and the difference in type is only due to the orientation of the observer. In this model we see a type 2 AGN if the BLR and accretion disc are blocked from our view by obscuring dust surrounding the AGN perpendicular to the axis of symmetry \citep{Keel1980}. However, orientation cannot change on the time-scale of the observed type changes, and hence some other explanation is needed.

We have shown using spectroscopy and multiwavelength photometry that NGC~1566 is in a high state with strong broad emission lines. The duration of the high state and the continuing variability are not consistent with potential explanations where the type change is due to tidal disruption or a once-off event such as a supernova. The recurrent brightening with CL events seen in NGC~1566 is probably common in CL AGNs, as such behaviour has also been noted for some other well-known CL AGNs such as NGC~4151 \citep{Oknyansky2016d}, NGC~5548 \citep{Bon2016} and NGC~2617 \citep{Oknyansky2017A, Oknyansky2018A}.

We propose that the change of type may be the result of increased luminosity causing the sublimation of dust in the line of sight which previously obscured part of the broad line region. This leads to a much more direct view of the central regions. The greater luminosity would also increase the intensity of the Balmer lines as well as highly ionized lines like [FeX] and [FeVII].

What is the reason for these recurrent outbursts in the object? The main problem with invoking a TDE (tidal disruption event) or supernova near SMBH (supermassive black hole) in some AGN is that they are extremely rare and cannot explain the comparatively greater rate of CL cases. Repeat tidal stars stripping \citep{Ivanov-Chernyakova06, Campana15} could lead to more frequent events \citep{Komossa17}, but this possibility is not sufficiently investigated yet. Understanding the physical process remains elusive at this stage.

\section{Summary}

We have shown, using spectroscopy (1.9 m SAAO) and multi-wavelength photometry (MASTER, {\it Swift} Ultraviolet/Optical and  XRT Telescopes), that NGC~1566 has just experienced a dramatic outburst in all wavelengths, including a considerable strengthening of broad permitted and high-ionization [FeX]6374\AA~lines, as well as substantial changes in the shape of the optical and X-ray continua. These confirm a new CL case for NGC~1566 where the AGN achieved brightness levels comparable to historical outbursts in about 1966 and 1992. We suspect that these strong outbursts may be recurrent events with a quasi-period of about 26 years. One possible interpretation for these outbursts involves tidal star stripping. More work is needed to determine the plausibility of this and alternative mechanisms.

\section*{Acknowledgements}
HW and FVW thank the SAAO for the generous allocation of telescope time which also resulted in the spectrum presented in this paper. We also express our thanks to the {\it Swift} ToO team for organizing and executing the observations. This work was supported in part by the Russian Foundation for Basic Research through grant 17-52-80139 BRICS-a and by the BRICS Multilateral Joint Science and Technology Research Collaboration grant 110480. MASTER work was supported by Lomonosov Moscow State University Development Programme and RSF grant 16-12-00085. DB is supported by the National Research Foundation of South Africa. We are grateful to S.~Komossa for useful discussions. 



\bibliographystyle{mnras}
\expandafter\ifx\csname natexlab\endcsname\relax\def\natexlab#1{#1}\fi

\interlinepenalty=10000

\bibliography{1566} 

\begin{thebibliography}{}
\makeatletter
\relax
\def\mn@urlcharsother{\let\do\@makeother \do\$\do\&\do\#\do\^\do\_\do\%\do\~}
\def\mn@doi{\begingroup\mn@urlcharsother \@ifnextchar [ {\mn@doi@}
  {\mn@doi@[]}}
\def\mn@doi@[#1]#2{\def\@tempa{#1}\ifx\@tempa\@empty \href
  {http://dx.doi.org/#2} {doi:#2}\else \href {http://dx.doi.org/#2} {#1}\fi
  \endgroup}
\def\mn@eprint#1#2{\mn@eprint@#1:#2::\@nil}
\def\mn@eprint@arXiv#1{\href {http://arxiv.org/abs/#1} {{\tt arXiv:#1}}}
\def\mn@eprint@dblp#1{\href {http://dblp.uni-trier.de/rec/bibtex/#1.xml}
  {dblp:#1}}
\def\mn@eprint@#1:#2:#3:#4\@nil{\def\@tempa {#1}\def\@tempb {#2}\def\@tempc
  {#3}\ifx \@tempc \@empty \let \@tempc \@tempb \let \@tempb \@tempa \fi \ifx
  \@tempb \@empty \def\@tempb {arXiv}\fi \@ifundefined
  {mn@eprint@\@tempb}{\@tempb:\@tempc}{\expandafter \expandafter \csname
  mn@eprint@\@tempb\endcsname \expandafter{\@tempc}}}

\bibitem[\protect\citeauthoryear{{Alloin}, {Pelat}, {Phillips}  \&
  {Whittle}}{{Alloin} et~al.}{1985}]{Alloin1985}
{Alloin} D.,  {Pelat} D.,  {Phillips} M.,   {Whittle} M.,  1985, \mn@doi [\apj]
  {10.1086/162783}, \href {http://adsabs.harvard.edu/abs/1985ApJ...288..205A}
  {288, 205}

\bibitem[\protect\citeauthoryear{{Alloin}, {Pelat}, {Phillips}, {Fosbury}  \&
  {Freeman}}{{Alloin} et~al.}{1986}]{Alloin1986}
{Alloin} D.,  {Pelat} D.,  {Phillips} M.~M.,  {Fosbury} R.~A.~E.,   {Freeman}
  K.,  1986, \mn@doi [\apj] {10.1086/164475}, \href
  {http://adsabs.harvard.edu/abs/1986ApJ...308...23A} {308, 23}

\bibitem[\protect\citeauthoryear{{Baribaud}, {Alloin}, {Glass}  \&
  {Pelat}}{{Baribaud} et~al.}{1992}]{Baribaud1992}
{Baribaud} T.,  {Alloin} D.,  {Glass} I.,   {Pelat} D.,  1992, \aap, \href
  {http://adsabs.harvard.edu/abs/1992A%26A...256..375B} {256, 375}

\bibitem[\protect\citeauthoryear{{Bon} et~al.,}{{Bon} et~al.}{2016}]{Bon2016}
{Bon} E.,  et~al., 2016, \mn@doi [\apjs] {10.3847/0067-0049/225/2/29}, \href
  {http://adsabs.harvard.edu/abs/2016ApJS..225...29B} {225, 29}

\bibitem[\protect\citeauthoryear{{Burrows} et~al.,}{{Burrows}
  et~al.}{2005}]{Burrows2005}
{Burrows} D.~N.,  et~al., 2005, \mn@doi [\ssr] {10.1007/s11214-005-5097-2},
  \href {http://adsabs.harvard.edu/abs/2005SSRv..120..165B} {120, 165}

\bibitem[\protect\citeauthoryear{{Campana}, {Mainetti}, {Colpi}, {Lodato},
  {D'Avanzo}, {Evans}  \& {Moretti}}{{Campana} et~al.}{2015}]{Campana15}
{Campana} S.,  {Mainetti} D.,  {Colpi} M.,  {Lodato} G.,  {D'Avanzo} P.,
  {Evans} P.~A.,   {Moretti} A.,  2015, \mn@doi [\aap]
  {10.1051/0004-6361/201525965}, \href
  {http://adsabs.harvard.edu/abs/2015A%26A...581A..17C} {581, A17}

\bibitem[\protect\citeauthoryear{{Clavel} et~al.,}{{Clavel}
  et~al.}{2000}]{Clavel2000}
{Clavel} J.,  et~al., 2000, \aap, \href
  {http://adsabs.harvard.edu/abs/2000A%26A...357..839C} {357, 839}

\bibitem[\protect\citeauthoryear{{Cutri}, {Mainzer}, {Dyk}  \& {Jiang}}{{Cutri}
  et~al.}{2018}]{Cutri2018}
{Cutri} R.~M.,  {Mainzer} A.~K.,  {Dyk} S.~D.~V.,   {Jiang} N.,  2018, Astron.
  Telegram, \href {http://adsabs.harvard.edu/abs/2018ATel11913....1C} {11913}

\bibitem[\protect\citeauthoryear{{Dai}, {Stanek}, {Kochanek}, {Shappee}  \&
  {ASAS-SN Collaboration}}{{Dai} et~al.}{2018}]{Dai2018}
{Dai} X.,  {Stanek} K.~Z.,  {Kochanek} C.~S.,  {Shappee} B.~J.,   {ASAS-SN
  Collaboration} 2018, Astron. Telegram, \href
  {http://adsabs.harvard.edu/abs/2018ATel11893....1D} {11893}

\bibitem[\protect\citeauthoryear{{Ducci}, {Siegert}, {Diehl},
  {Sanchez-Fernandez}, {Ferrigno}, {Savchenko}  \& {Bozzo}}{{Ducci}
  et~al.}{2018}]{Ducci2018}
{Ducci} L.,  {Siegert} T.,  {Diehl} R.,  {Sanchez-Fernandez} C.,  {Ferrigno}
  C.,  {Savchenko} V.,   {Bozzo} E.,  2018, Astron. Telegram, \href
  {http://adsabs.harvard.edu/abs/2018ATel11754....1D} {11754}

\bibitem[\protect\citeauthoryear{{Evans} et~al.,}{{Evans}
  et~al.}{2009}]{Evans2009}
{Evans} P.~A.,  et~al., 2009, \mn@doi [\mnras]
  {10.1111/j.1365-2966.2009.14913.x}, \href
  {http://adsabs.harvard.edu/abs/2009MNRAS.397.1177E} {397, 1177}

\bibitem[\protect\citeauthoryear{{Ferrigno}, {Siegert}, {Sanchez-Fernandez},
  {Kuulkers}, {Ducci}, {Savchenko}  \& {Bozzo}}{{Ferrigno}
  et~al.}{2018}]{Ferrigno2018}
{Ferrigno} C.,  {Siegert} T.,  {Sanchez-Fernandez} C.,  {Kuulkers} E.,  {Ducci}
  L.,  {Savchenko} V.,   {Bozzo} E.,  2018, Astron. Telegram, \href
  {http://adsabs.harvard.edu/abs/2018ATel11783....1F} {11783}

\bibitem[\protect\citeauthoryear{{Gehrels} et~al.,}{{Gehrels}
  et~al.}{2004}]{Gehrels2004}
{Gehrels} N.,  et~al., 2004, \mn@doi [\apj] {10.1086/422091}, \href
  {http://adsabs.harvard.edu/abs/2004ApJ...611.1005G} {611, 1005}

\bibitem[\protect\citeauthoryear{{Grupe}, {Komossa}  \& {Schartel}}{{Grupe}
  et~al.}{2018}]{Grupe2018}
{Grupe} D.,  {Komossa} S.,   {Schartel} N.,  2018, Astron. Telegram, \href
  {http://adsabs.harvard.edu/abs/2018ATel11903....1G} {11903}

\bibitem[\protect\citeauthoryear{{Hamuy}, {Suntzeff}, {Heathcote}, {Walker},
  {Gigoux}  \& {Phillips}}{{Hamuy} et~al.}{1994}]{Hamuy1994}
{Hamuy} M.,  {Suntzeff} N.~B.,  {Heathcote} S.~R.,  {Walker} A.~R.,  {Gigoux}
  P.,   {Phillips} M.~M.,  1994, \mn@doi [\pasp] {10.1086/133417}, \href
  {http://adsabs.harvard.edu/abs/1994PASP..106..566H} {106, 566}

\bibitem[\protect\citeauthoryear{{Ivanov} \& {Chernyakova}}{{Ivanov} \&
  {Chernyakova}}{2006}]{Ivanov-Chernyakova06}
{Ivanov} P.~B.,  {Chernyakova} M.~A.,  2006, \mn@doi [\aap]
  {10.1051/0004-6361:20053409}, \href
  {http://adsabs.harvard.edu/abs/2006A%26A...448..843I} {448, 843}

\bibitem[\protect\citeauthoryear{{Kawamuro}, {Ueda}, {Tazaki}  \&
  {Terashima}}{{Kawamuro} et~al.}{2013}]{Kawamuro2013}
{Kawamuro} T.,  {Ueda} Y.,  {Tazaki} F.,   {Terashima} Y.,  2013, \mn@doi
  [\apj] {10.1088/0004-637X/770/2/157}, \href
  {http://adsabs.harvard.edu/abs/2013ApJ...770..157K} {770, 157}

\bibitem[\protect\citeauthoryear{{Keel}}{{Keel}}{1980}]{Keel1980}
{Keel} W.~C.,  1980, \mn@doi [\aj] {10.1086/112662}, \href
  {http://adsabs.harvard.edu/abs/1980AJ.....85..198K} {85, 198}

\bibitem[\protect\citeauthoryear{{Kinney}, {Calzetti}, {Bohlin}, {McQuade},
  {Storchi-Bergmann}  \& {Schmitt}}{{Kinney} et~al.}{1996}]{Kinney1996}
{Kinney} A.~L.,  {Calzetti} D.,  {Bohlin} R.~C.,  {McQuade} K.,
  {Storchi-Bergmann} T.,   {Schmitt} H.~R.,  1996, \mn@doi [\apj]
  {10.1086/177583}, \href {http://adsabs.harvard.edu/abs/1996ApJ...467...38K}
  {467, 38}

\bibitem[\protect\citeauthoryear{{Kochanek} et~al.,}{{Kochanek}
  et~al.}{2017}]{Kochanek2017}
{Kochanek} C.~S.,  et~al., 2017, \mn@doi [\pasp] {10.1088/1538-3873/aa80d9},
  \href {http://adsabs.harvard.edu/abs/2017PASP..129j4502K} {129, 104502}

\bibitem[\protect\citeauthoryear{{Komossa} et~al.,}{{Komossa}
  et~al.}{2017}]{Komossa17}
{Komossa} S.,  et~al., 2017, in {Gomboc} A.,  ed.,  IAU Symposium Vol. 324, New
  Frontiers in Black Hole Astrophysics. pp 168--171

\bibitem[\protect\citeauthoryear{{Kornilov} et~al.,}{{Kornilov}
  et~al.}{2012}]{Kornilov2012}
{Kornilov} V.~G.,  et~al., 2012, \mn@doi [Experimental Astronomy]
  {10.1007/s10686-011-9280-z}, \href
  {http://adsabs.harvard.edu/abs/2012ExA....33..173K} {33, 173}

\bibitem[\protect\citeauthoryear{{Kriss}, {Hartig}, {Armus}, {Blair},
  {Caganoff}  \& {Dressel}}{{Kriss} et~al.}{1991}]{Kriss1991}
{Kriss} G.~A.,  {Hartig} G.~F.,  {Armus} L.,  {Blair} W.~P.,  {Caganoff} S.,
  {Dressel} L.,  1991, \mn@doi [\apjl] {10.1086/186105}, \href
  {http://adsabs.harvard.edu/abs/1991ApJ...377L..13K} {377, L13}

\bibitem[\protect\citeauthoryear{{Kuin}, {Bozzo}, {Ferrigno}, {Savchenko},
  {Kuulkers}, {Ducci}  \& {Ducci}}{{Kuin} et~al.}{2018}]{Kuin2018}
{Kuin} P.,  {Bozzo} E.,  {Ferrigno} C.,  {Savchenko} V.,  {Kuulkers} E.,
  {Ducci} C.~S.-F.~L.,   {Ducci} L.,  2018, Astron. Telegram, \href
  {http://adsabs.harvard.edu/abs/2018ATel11786....1K} {11786}

\bibitem[\protect\citeauthoryear{{Landt}, {Ward}, {Steenbrugge}  \&
  {Ferland}}{{Landt} et~al.}{2015a}]{Landt2015}
{Landt} H.,  {Ward} M.~J.,  {Steenbrugge} K.~C.,   {Ferland} G.~J.,  2015a,
  \mn@doi [\mnras] {10.1093/mnras/stv062}, \href
  {http://adsabs.harvard.edu/abs/2015MNRAS.449.3795L} {449, 3795}

\bibitem[\protect\citeauthoryear{{Landt}, {Ward}, {Steenbrugge}  \&
  {Ferland}}{{Landt} et~al.}{2015b}]{Landt2015b}
{Landt} H.,  {Ward} M.~J.,  {Steenbrugge} K.~C.,   {Ferland} G.~J.,  2015b,
  \mn@doi [\mnras] {10.1093/mnras/stv2176}, \href
  {http://adsabs.harvard.edu/abs/2015MNRAS.454.3688L} {454, 3688}

\bibitem[\protect\citeauthoryear{{Lipunov} et~al.,}{{Lipunov}
  et~al.}{2010}]{Lipunov2010}
{Lipunov} V.,  et~al., 2010, \mn@doi [Adv. Astron.] {10.1155/2010/349171},
  \href {http://adsabs.harvard.edu/abs/2010AdAst2010E..30L} {2010, 349171}

\bibitem[\protect\citeauthoryear{{Netzer}}{{Netzer}}{1974}]{Netzer1974}
{Netzer} H.,  1974, \mn@doi [\mnras] {10.1093/mnras/169.3.579}, \href
  {http://adsabs.harvard.edu/abs/1974MNRAS.169..579N} {169, 579}

\bibitem[\protect\citeauthoryear{{Netzer}}{{Netzer}}{2015}]{Netzer2015}
{Netzer} H.,  2015, \mn@doi [\araa] {10.1146/annurev-astro-082214-122302},
  \href {http://adsabs.harvard.edu/abs/2015ARA%26A..53..365N} {53, 365}

\bibitem[\protect\citeauthoryear{{Oknyanskii}, {Lyutyi}  \&
  {Chuvaev}}{{Oknyanskii} et~al.}{1991}]{Oknyansky1991}
{Oknyanskii} V.~L.,  {Lyutyi} V.~M.,   {Chuvaev} K.~K.,  1991, Soviet Astronomy
  Letters, \href {http://adsabs.harvard.edu/abs/1991SvAL...17..100O} {17, 100}

\bibitem[\protect\citeauthoryear{{Oknyanskij} \& {Chuvaev}}{{Oknyanskij} \&
  {Chuvaev}}{1982}]{Oknyansky1982}
{Oknyanskij} V.~L.,  {Chuvaev} K.~K.,  1982, Astronomicheskij Tsirkulyar, \href
  {http://adsabs.harvard.edu/abs/1982ATsir1228....1O} {1228, 1}

\bibitem[\protect\citeauthoryear{{Oknyanskij} \& {Horne}}{{Oknyanskij} \&
  {Horne}}{2001}]{Oknyansky2001}
{Oknyanskij} V.~L.,  {Horne} K.,  2001, in {Peterson} B.~M.,  {Pogge} R.~W.,
  {Polidan} R.~S.,  eds,  ASP Conf. Ser. Vol. 224, Probing the Physics of
  Active Galactic Nuclei. Astron. Soc. Pac. San Francisco. p.~149

\bibitem[\protect\citeauthoryear{{Oknyanskij}, {Metlova}, {Huseynov}, {Guo}  \&
  {Lyuty}}{{Oknyanskij} et~al.}{2016}]{Oknyansky2016d}
{Oknyanskij} V.~L.,  {Metlova} N.~V.,  {Huseynov} N.~A.,  {Guo} D.-F.,
  {Lyuty} V.~M.,  2016, \mn@doi [Odessa Astronomical Publications]
  {10.18524/1810-4215.2016.29.85058}, \href
  {http://adsabs.harvard.edu/abs/2016OAP....29...95O} {29, 95}

\bibitem[\protect\citeauthoryear{{Oknyansky} et~al.,}{{Oknyansky}
  et~al.}{2017}]{Oknyansky2017A}
{Oknyansky} V.~L.,  et~al., 2017, \mn@doi [\mnras] {10.1093/mnras/stx149},
  \href {http://adsabs.harvard.edu/abs/2017MNRAS.467.1496O} {467, 1496}

\bibitem[\protect\citeauthoryear{{Oknyansky} et~al.,}{{Oknyansky}
  et~al.}{2018a}]{Oknyansky2018A}
{Oknyansky} V.,  et~al., 2018a, Astron. Telegram, \href
  {http://adsabs.harvard.edu/abs/2018ATel11703....1O} {11703}

\bibitem[\protect\citeauthoryear{{Oknyansky}, {Lipunov}, {Gorbovskoy},
  {Winkler}, {van Wyk}, {Tsygankov}  \& {Buckley}}{{Oknyansky}
  et~al.}{2018b}]{Oknyansky2018B}
{Oknyansky} V.~L.,  {Lipunov} V.~M.,  {Gorbovskoy} E.~S.,  {Winkler} H.,  {van
  Wyk} F.,  {Tsygankov} S.,   {Buckley} D.~A.~H.,  2018b, Astron. Telegram,
  \href {http://adsabs.harvard.edu/abs/2018ATel11915....1O} {11915}

\bibitem[\protect\citeauthoryear{{Osmer}, {Smith}  \& {Weedman}}{{Osmer}
  et~al.}{1974}]{Osmer1974}
{Osmer} P.~S.,  {Smith} M.~G.,   {Weedman} D.~W.,  1974, \mn@doi [\apj]
  {10.1086/152787}, \href {http://adsabs.harvard.edu/abs/1974ApJ...189..187O}
  {189, 187}

\bibitem[\protect\citeauthoryear{{Osterbrock} \& {Shuder}}{{Osterbrock} \&
  {Shuder}}{1982}]{Osterbrock1982}
{Osterbrock} D.~E.,  {Shuder} J.~M.,  1982, \mn@doi [\apjs] {10.1086/190793},
  \href {http://adsabs.harvard.edu/abs/1982ApJS...49..149O} {49, 149}

\bibitem[\protect\citeauthoryear{{Parker} et~al.,}{{Parker}
  et~al.}{2016}]{Parker2016}
{Parker} M.~L.,  et~al., 2016, \mn@doi [\mnras] {10.1093/mnras/stw1449}, \href
  {http://adsabs.harvard.edu/abs/2016MNRAS.461.1927P} {461, 1927}

\bibitem[\protect\citeauthoryear{{Pastoriza} \& {Gerola}}{{Pastoriza} \&
  {Gerola}}{1970}]{Pastoriza1970}
{Pastoriza} M.,  {Gerola} H.,  1970, \aplett, \href
  {http://adsabs.harvard.edu/abs/1970ApL.....6..155P} {6, 155}

\bibitem[\protect\citeauthoryear{{Peterson}}{{Peterson}}{1988}]{Peterson1988}
{Peterson} B.~M.,  1988, \mn@doi [\pasp] {10.1086/132130}, \href
  {http://adsabs.harvard.edu/abs/1988PASP..100...18P} {100, 18}

\bibitem[\protect\citeauthoryear{{Ricci} et~al.,}{{Ricci}
  et~al.}{2016}]{Ricci2016}
{Ricci} C.,  et~al., 2016, \mn@doi [\apj] {10.3847/0004-637X/820/1/5}, \href
  {http://adsabs.harvard.edu/abs/2016ApJ...820....5R} {820, 5}

\bibitem[\protect\citeauthoryear{{Rose}, {Elvis}  \& {Tadhunter}}{{Rose}
  et~al.}{2015}]{Rose2015}
{Rose} M.,  {Elvis} M.,   {Tadhunter} C.~N.,  2015, \mn@doi [\mnras]
  {10.1093/mnras/stv113}, \href
  {http://adsabs.harvard.edu/abs/2015MNRAS.448.2900R} {448, 2900}

\bibitem[\protect\citeauthoryear{{Shappee} et~al.,}{{Shappee}
  et~al.}{2014}]{Shappee2014}
{Shappee} B.~J.,  et~al., 2014, \mn@doi [\apj] {10.1088/0004-637X/788/1/48},
  \href {http://adsabs.harvard.edu/abs/2014ApJ...788...48S} {788, 48}

\bibitem[\protect\citeauthoryear{{Shobbrook}}{{Shobbrook}}{1966}]{Shobbrook1966}
{Shobbrook} R.~R.,  1966, \mn@doi [\mnras] {10.1093/mnras/131.3.365}, \href
  {http://adsabs.harvard.edu/abs/1966MNRAS.131..365S} {131, 365}

\bibitem[\protect\citeauthoryear{{Storey} \& {Zeippen}}{{Storey} \&
  {Zeippen}}{2000}]{Storey2000}
{Storey} P.~J.,  {Zeippen} C.~J.,  2000, \mn@doi [\mnras]
  {10.1046/j.1365-8711.2000.03184.x}, \href
  {http://adsabs.harvard.edu/abs/2000MNRAS.312..813S} {312, 813}

\bibitem[\protect\citeauthoryear{{Ulrich}, {Maraschi}  \& {Urry}}{{Ulrich}
  et~al.}{1997}]{Ulrich1997}
{Ulrich} M.-H.,  {Maraschi} L.,   {Urry} C.~M.,  1997, \mn@doi [\araa]
  {10.1146/annurev.astro.35.1.445}, \href
  {http://adsabs.harvard.edu/abs/1997ARA%26A..35..445U} {35, 445}

\bibitem[\protect\citeauthoryear{{Veilleux}}{{Veilleux}}{1988}]{Veilleux1988}
{Veilleux} S.,  1988, \mn@doi [\aj] {10.1086/114766}, \href
  {http://adsabs.harvard.edu/abs/1988AJ.....95.1695V} {95, 1695}

\bibitem[\protect\citeauthoryear{{Wachter}, {Leach}  \& {Kellogg}}{{Wachter}
  et~al.}{1979}]{Wachter1979}
{Wachter} K.,  {Leach} R.,   {Kellogg} E.,  1979, \mn@doi [\apj]
  {10.1086/157084}, \href {http://adsabs.harvard.edu/abs/1979ApJ...230..274W}
  {230, 274}

\bibitem[\protect\citeauthoryear{{Winkler}}{{Winkler}}{1992}]{Winkler1992}
{Winkler} H.,  1992, \mn@doi [\mnras] {10.1093/mnras/257.4.677}, \href
  {http://adsabs.harvard.edu/abs/1992MNRAS.257..677W} {257, 677}

\bibitem[\protect\citeauthoryear{{da Silva}, {Steiner}  \& {Menezes}}{{da
  Silva} et~al.}{2017}]{Silva2017}
{da Silva} P.,  {Steiner} J.~E.,   {Menezes} R.~B.,  2017, \mn@doi [\mnras]
  {10.1093/mnras/stx1458}, \href
  {http://adsabs.harvard.edu/abs/2017MNRAS.470.3850D} {470, 3850}

\bibitem[\protect\citeauthoryear{{de Vaucouleurs} \& {de Vaucouleurs}}{{de
  Vaucouleurs} \& {de Vaucouleurs}}{1961}]{Vaucouleurs1961}
{de Vaucouleurs} G.,  {de Vaucouleurs} A.,  1961, \memras, \href
  {http://adsabs.harvard.edu/abs/1961MmRAS..68...69D} {68, 69}

\makeatother
\end{thebibliography}

\bsp	

\label{lastpage}
\end{document}